\documentclass[aps,pre]{revtex4}
\usepackage{graphicx}
\begin{document}
\title{
Beyond the poor man's implementation of unconditionally stable algorithms\\
to solve the time-dependent Maxwell Equations}
\author{J.S. Kole\footnote{E-mail: j.s.kole@phys.rug.nl},
        M.T. Figge\footnote{E-mail: m.t.figge@phys.rug.nl},
    and H. De Raedt\footnote{E-mail: h.a.deraedt@phys.rug.nl\\
                                     http://rugth30.phys.rug.nl/compphys}}
\affiliation{
Centre for Theoretical Physics and Materials Science Centre\\
University of Groningen, Nijenborgh 4\\
NL-9747 AG Groningen, The Netherlands
}
\date{\today}

%%%%%%%%%%%%%%%%%%%%%%%%%%%%%%%%%%%%%%%%%%%%%%%%%%%%%%%%%%%%%%%%%%%%%%%%%%%%%

\begin{abstract}

\bigskip

For the recently introduced algorithms to solve the time-dependent Maxwell
equations~\cite{Kole01}, we construct a variable grid implementation and an
improved spatial discretization implementation that preserve the exceptional
property of the algorithms to be unconditionally stable by construction.
We find that the performance and accuracy of the corresponding algorithms are
significant and illustrate their practical relevance by simulating various
physical model systems.

\end{abstract}

\maketitle

%%%%%%%%%%%%%%%%%%%%%%%%%%%%%%%%%%%%%%%%%%%%%%%%%%%
% MACROS
%
\newcommand{\dd}[1]{\frac{\partial}{\partial #1}}
\newcommand{\ve}{\varepsilon}
\def\be{{\mathbf{e}}}
\def\bH{{\mathbf{H}}}
\def\bE{{\mathbf{E}}}
\def\br{{\mathbf{r}}}
\def\bX{{\mathbf{X}}}
\def\bY{{\mathbf{Y}}}
\newcommand{\php}{\phantom{+}}
\newcommand{\pht}{\phantom{T}}
%
%\baselineskip=28pt
%
%
%%%%%%%%%%%%%%%%%%%%%%%%%%%%%%%%%%%%%%%%%%%%%%%%%%%%%%%%%%%%%%%%%%%%%%%%%%%%
%
\section{Introduction}\label{sec1}
%
%%%%%%%%%%%%%%%%%%%%%%%%%%%%%%%%%%%%%%%%%%%%%%%%%%%%%%%%%%%%%%%%%%%%%%%%%%%%
%

In a recent paper, we introduced a family of algorithms to solve the
time-dependent Maxwell equations (TDME)~\cite{Kole01}.
Salient features of these algorithms include the rigorously provable
unconditional stability for $d$-dimensional systems ($d=1,2,3$) with
spatially varying permittivity and permeability, as well as the exact
conservation of the energy density of the electromagnetic (EM) fields.
Furthermore, we have demonstrated that - without affecting the unconditional
stability of the algorithms - the order of accuracy in the time integration
can be systematically increased.
An important aspect that has not been considered in our earlier
work~\cite{Kole01} concerns the effect of the discretization of space on the
accuracy of the algorithms.
Previously, we employed only the most simple spatial discretization, namely
the central-difference scheme on a cartesian grid with a constant mesh
size~\cite{Kole01}.
We refer to this basic spatial discretization scheme as the {\it poor man's
implementation}.
Many numerical simulations of realistic physical systems require algorithms
with a more accurate spatial discretization and a more flexible spatial grid
for an optimal use of computer resources (CPU time and computer memory).
In the present paper we show that implementing a fourth-order accurate
approximation of the spatial derivatives and a spatial grid of variable
mesh sizes preserve the unconditional stability of the algorithms.
We simulate various physical model systems using these new implementations to
demonstrate the significant improvement with respect to the required computer
resources in the computation of eigenmode spectra and to study systematically
the temporal and spatial accuracy of the algorithms.

Our presentation is organized as follows:
We recapitulate the theory of constructing unconditionally stable algorithms
to solve the TDME in Sec.~\ref{sec2} and describe the basic properties of the
poor man's implementation in Sec.~\ref{sec3}.
Then, in Sec.~\ref{sec4} and Sec.~\ref{sec5}, we present the implementation of,
respectively, the variable grid and the improved spatial discretization.
Our conclusions are given in Sec.~\ref{sec6}.

%
%-------------------------------------------------------------
%
\section{Unconditionally stable algorithms to solve
Maxwell's equations}\label{sec2}
%
%-------------------------------------------------------------
%
We consider a $d$-dimensional model system of EM fields in a medium with
spatially varying permittivity and/or permeability, surrounded by a perfectly
conducting box.
In the absence of free charges and currents, the EM fields in such a system
satisfy Maxwell's equations~\cite{BornWolf}
\begin{eqnarray}
\dd{t}\,\bH  =  -\frac{1}{\mu} \nabla \times \bE\;\;\;\; & {\rm and}& \;\;\;\;
\dd{t}\,{\bE} = \frac{1}{\ve} \nabla \times {\bH}\,,
\label{max}\\[0.3cm]
{\rm div}\,\ve\bE  =  0\;\;\;\; & {\rm and} & \;\;\;\;
{\rm div}\,\bH  = 0\,,
\label{maxdiv}\\[-0.3cm]\nonumber
\end{eqnarray}
where $\bH=(H_x(\br,t),H_y(\br,t),H_z(\br,t))^T$ and
$\bE=(E_x(\br,t),E_y(\br,t),E_z(\br,t))^T$ denote, respectively, the magnetic
and the electric field vector.
The permeability and the permittivity are given
by $\mu=\mu(\br)$ and $\ve=\ve(\br)$.
For simplicity of notation, we will omit the spatial dependence on
$\br=(x,y,z)^T$ unless this leads to ambiguities.
On the surface of the perfectly conducting box the EM fields satisfy
the boundary conditions~\cite{BornWolf}
\begin{eqnarray}
\mathbf{n} \times \bE = \mathbf{0} \;\;\;\;{\rm and}\;\;\;\;
\mathbf{n} \cdot \bH = {\rm 0}\,, \label{boundcond}
\end{eqnarray}
with $\mathbf{n}$ denoting the vector normal to a boundary of the
surface.
The conditions Eqs.~(\ref{boundcond}) assure that the normal component
of the magnetic field and the tangential components of the electric
field vanish at the boundary~\cite{BornWolf}.
Some important symmetries of the Maxwell equations
(\ref{max})-(\ref{maxdiv}) can be made explicit by introducing the
fields
\begin{eqnarray}
\bX(t)=\sqrt{\mu}\bH(t)\;\;\;\;{\rm and}\;\;\;\;
\bY(t)=\sqrt{\varepsilon}\bE(t)\,. \label{EY}
\end{eqnarray}
In terms of the fields $\bX(t)$ and $\bY(t)$, the TDME (\ref{max}) read
\begin{equation} \dd{t} \left(\begin{array}{c} \bX(t) \\
\bY(t) \end{array} \right) =
 \left( \begin{array}{cc} 0 &
  -\frac{1}{\sqrt{\mu}}\mathbf{\nabla}\times\frac{1}
  {\sqrt{\varepsilon}} \\
   \frac{1}{\sqrt{\varepsilon}}\mathbf{\nabla}
   \times\frac{1}{\sqrt{\mu}} & 0
 \end{array}\right)
 \left(\begin{array}{c} \bX(t) \\ \bY(t)  \end{array} \right)
 \equiv {\mathcal H}
 \left(\begin{array}{c} \bX(t) \\ \bY(t)  \end{array} \right)\,.
\label{eqn:mtxeqn}
\end{equation}
Writing $\mathbf{\Psi}(t)=(\bX(t),\bY(t))^T$, Eq.~(\ref{eqn:mtxeqn}) becomes
\begin{equation}
\dd{t}\mathbf{\Psi}(t)={\mathcal H}\mathbf{\Psi}(t)\,. \label{TDME}
\end{equation}
It is easy to show that ${\mathcal H}$ is skew-symmetric, i.e.
${\mathcal H}^T=-{\mathcal H}$, with respect to the inner product
$\langle\mathbf{\Psi}|\mathbf{\Psi}^\prime\rangle\equiv\int_V\mathbf{\Psi}^T\cdot\mathbf{\Psi}^\prime\, d\br$,
where $V$ denotes the volume of the enclosing box.
The formal solution of Eq.~(\ref{TDME}) is given by
\begin{equation}
\mathbf{\Psi}(t) = U(t)\mathbf{\Psi}(0) = e^{t{\mathcal H}}\mathbf{\Psi}(0)\,,
\end{equation}
where $\mathbf{\Psi}(0)$ represents the initial state of the EM fields.
The operator $U(t)= e^{t{\mathcal H}}$ determines the time
evolution.
By construction
$\|\mathbf{\Psi}(t)\|^2 = \langle \mathbf{\Psi}(t)|\mathbf{\Psi}(t)\rangle =\int_V
\left[\varepsilon{\bE}^2(t) +\mu{\bH}^2(t)
\right] \, d{\br}$,
relating the length of $\mathbf{\Psi}(t)$ to the energy density
$w(t)\equiv
\varepsilon{\bE}^2(t)+\mu{\bH}^2(t)$
of the EM fields~\cite{BornWolf}.
As $U(t)^T=U(-t)=U^{-1}(t)=e^{-t{\mathcal H}}$ it follows that
$\langle U(t)\mathbf{\Psi}(0)|U(t)\mathbf{\Psi}(0)\rangle=\langle\mathbf{\Psi}(t)|\mathbf{\Psi}(t)\rangle=
\langle\mathbf{\Psi}(0)|\mathbf{\Psi}(0)\rangle$.
Hence the time-evolution operator $U(t)$ is an orthogonal transformation,
rotating the vector $\mathbf{\Psi}(t)$ without changing its length $\|\mathbf{\Psi}\|$.
In physical terms this means that the energy density of the EM fields
does not change with time, as expected on physical grounds~\cite{BornWolf}.

A numerical procedure that solves the TDME necessarily starts
by discretizing the spatial derivatives.
This maps the continuum problem described by ${\mathcal H}$ onto a
lattice problem defined by a matrix $H$.
Ideally, this mapping should not change the basic symmetries of the
original problem.
The underlying symmetry of the TDME suggests to use matrices $H$ that are
real and skew-symmetric.
Since formally the time evolution of the EM fields on the lattice is
given by
$\mathbf{\Psi}(t+\tau) = U(\tau)\mathbf{\Psi}(t) = e^{\tau{H}}\mathbf{\Psi}(t)$,
the second ingredient of the numerical procedure is to choose an
approximation of the time-evolution operator $U(\tau)$.
The fact that $U(t)$ is an orthogonal transformation is essential for
the development of an unconditionally stable algorithm to solve the
Maxwell equations~\cite{Kole01}.
A systematic approach to construct orthogonal approximations to matrix
exponentials
%, i.e. to construct unconditionally stable algorithms,
is to make use of the Lie-Trotter-Suzuki formula~\cite{Trotter59,Suzuki77}
\begin{equation}
e^{t(H_1+\ldots+H_p)}=
\lim_{m\rightarrow\infty}
\left(\prod_{i=1}^p e^{t{H}_i/m}\right)^m,
\label{TROT}
\end{equation}
and generalizations thereof~\cite{Suzuki8591,DeRaedt83}.
Applied to the case of interest here, the success of this approach relies
on the basic but rather trivial premise that the matrix $H$ can be written
as ${H}=\sum_{i=1}^{p}{H}_i$, where each of the matrices ${H}_i$ is real
and skew-symmetric.
Expression Eq.~(\ref{TROT}) suggests that
\begin{equation}
U_1(\tau)=e^{\tau{H}_1}\ldots e^{\tau{H}_p}
\label{tsapprox}
\end{equation}
might be a good approximation to $U(\tau)$ if $\tau$ is sufficiently small.
In fact, it can be shown that $U(\tau)$ and $U_1(\tau)$ are identical up to
first order in
$\tau$~\cite{DeRaedt87}.
%and we will call $U_1(\tau)$ a first-order approximation
%to $U(\tau)$
Most importantly, if all the $H_i$ are real and skew-symmetric, $U_1(\tau)$
is orthogonal by construction.
Therefore, by construction, a numerical scheme based on
Eq.~(\ref{tsapprox}) will be unconditionally stable.
The product-formula approach provides simple, systematic procedures to
improve the accuracy of the approximation to $U(\tau)$ without changing its
fundamental symmetries.
For example the orthogonal matrix
\begin{equation}
U_2(\tau)\;=\;U_1^T(-\tau/2) \,U_1(\tau/2)
\label{secordapp}
\end{equation}
is identical to $U(\tau)$ up to second order in
$\tau$~\cite{Suzuki8591,DeRaedt83}.
Suzuki's fractal decomposition approach~\cite{Suzuki8591} gives a general
method to construct higher-order approximations based on $U_1(\tau)$ or
$U_2(\tau)$.
A particularly useful approximation, which is identical to $U(\tau)$ up to
fourth order in $\tau$, is given by~\cite{Suzuki8591}
\begin{equation}
U_4(\tau)=U_2(a\tau)U_2(a\tau)U_2((1-4a)\tau)U_2(a\tau)U_2(a\tau)\,,
\label{fouordapp}
\end{equation}
where $a=1/(4-4^{1/3})$.
From Eqs.~(\ref{tsapprox})-(\ref{fouordapp}) it follows that, in practice,
an efficient implementation of a scheme based on $U_1(\tau)$ is all that is
needed to construct the higher-order algorithms Eqs.~(\ref{secordapp}) and
(\ref{fouordapp}).
In many applications the approximations $U_n(\tau)$ to the time-evolution
operator $U(t)$ have proven to be very
useful~\cite{Suzuki77,DeRaedt83,DeRaedt87,Koboyashi94,DeRaedt94,Rouhi95,Shadwick97,Krech98,Tran98,Michielsen98,DeRaedt00}
and turn out to be equally useful for solving the TDME~\cite{Kole01}.
In particular, it can be shown that the difference between the exact
EM field vector $\mathbf{\Psi}(t)=U(t)\mathbf{\Psi}(0)$ and the approximate
one, $\mathbf{\Psi}_n(t)=U_n(t)\mathbf{\Psi}(0)$, is bounded
by~\cite{DeRaedt87}
\begin{equation}
\|(U(t)-U_n(t))\mathbf{\Psi}(0)\|\;=\;
\|\mathbf{\Psi}(t)-\mathbf{\Psi}_n(t)\|\;\leq\;C_n\,t\,\tau^n\,,
\label{psierrortau}
\end{equation}
where $C_n$ is a constant.
The rigorous upper bound on the error of the EM field vector will be used to
specify unconditionally stable algorithms by the temporal and spatial accuracy
of the computed EM field.
We denote an algorithm by T$n$S$m$ if its implementation involves a time
integration based on $U_n(\tau)$ and a spatial discretization scheme based on
an $m$th-order accurate approximation of the spatial derivatives.

%
%-------------------------------------------------------------
%
\section{Poor man's implementation}\label{sec3}
%
%-------------------------------------------------------------
%

In this section, we briefly recapitulate the construction of the
unconditionally stable algorithm to solve Maxwell's equations in a
one-dimensional (1D) system.
Furthermore, we discuss general properties of this implementation refering
also to the two-dimensional (2D) and three-dimensional (3D) case.

Maxwell's equations for a 1D system extending along the $x$-axis
contain no partial derivatives with respect to $y$ or $z$.
Also $\ve$ and $\mu$ do not depend on $y$ or $z$.
Under these conditions, the TDME reduce to two independent sets
of first-order differential equations~\cite{BornWolf}.
The solutions to these sets are known as the transverse electric (TE) mode
and the transverse magnetic (TM) mode~\cite{BornWolf}.
%As the equations of the TE- and TM-mode only differ by a sign
Restricting our considerations to the TM-mode, it follows
from Eq.~(\ref{eqn:mtxeqn}) that the magnetic field
$H_y(x,t)=X_y(x,t)/\sqrt{\mu(x)}$ and the electric field
$E_z(x,t)=Y_z(x,t)/\sqrt{\ve(x)}$ are solutions of
\begin{eqnarray}
\dd{t}X_y(x,t) & = &
\frac{1}{\sqrt{\mu(x)}}\dd{x}\left(\frac{Y_z(x,t)}{\sqrt{\ve(x)}}\right)\,,\\
\dd{t}Y_z(x,t) & = &
\frac{1}{\sqrt{\ve(x)}}\dd{x}\left(\frac{X_y(x,t)}{\sqrt{\mu(x)}}\right)\,.
\end{eqnarray}
Note that in 1D the divergence of $H_y(x,t)$ and $E_z(x,t)$ is zero, hence
Eqs.~(\ref{maxdiv}) are automatically satisfied.
Using the central-difference scheme, which yields a second-order accurate
approximation of the spatial derivatives, we obtain
\begin{eqnarray}
\dd{t}X_y(i,t) & = &
\frac{1}{\delta\sqrt{\mu_i}} \left(
\frac{Y_z(i+1,t)}{\sqrt{\varepsilon_{i+1}}}-
\frac{Y_z(i-1,t)}{\sqrt{\varepsilon_{i-1}}}\right)\,,
\label{eqn:discrTMX} \\
\dd{t}Y_z(j,t) & = &
\frac{1}{\delta\sqrt{\varepsilon_{j}}} \left(
\frac{X_y(j+1,t)}{\sqrt{\mu_{j+1}}}-
\frac{X_y(j-1,t)}{\sqrt{\mu_{j-1}}}\right)\,,
\label{eqn:discrTMY}
\end{eqnarray}
where the spatial coordinate of an EM field component is specified through
the lattice index $i$, e.g. $X_y(i,t)$ stands for $X_y(x=i\delta/2,t)$, and
$\delta/2$ the distance between two neighboring lattice points.
\begin{figure}[h]
\begin{center}
\includegraphics[width=8.cm]{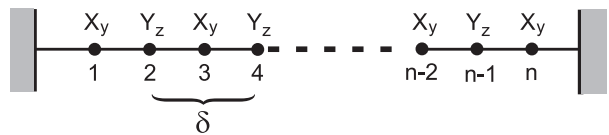}
\caption{Positions of the two TM-mode EM field components on the 1D grid.
\label{fig:fig1}}
\end{center}
\end{figure}
Following Yee~\cite{Yee66} it is convenient to assign $X_y(i,t)$ and
$Y_z(j,t)$ to the odd, respectively, even numbered lattice site, as shown
in Fig.~\ref{fig:fig1} for a grid of $n$ points.
The equations (\ref{eqn:discrTMX}) and (\ref{eqn:discrTMY}) can now be
combined into one equation of the form Eq.~(\ref{TDME}) by introducing
the $n$-dimensional vector $\mathbf{\Psi}(t)$ with elements
\begin{equation}
\Psi(i,t) =
\left\{ \begin{array}{lll} X_y(i,t)=\sqrt{\mu_i}H_y(i,t), &\;\;\; \mbox{$i$ odd}
\\ Y_z(i,t)=\sqrt{\varepsilon_i}E_z(i,t), &\;\;\; \mbox{$i$ even}
\end{array} \right..
\label{eqn:eqnindx}
\end{equation}
The vector $\mathbf{\Psi}(t)$ describes both the magnetic and the electric field
on the lattice points $i=1,\ldots,n$ and the $i$th element of $\mathbf{\Psi}(t)$
is given by the inner product $\Psi(i,t)=\be^T_i\cdot\mathbf{\Psi}(t)$, where
$\be_i$ denotes the $i$th unit vector in the $n$-dimensional vector
space.
Using this notation, it is easy to show that
\begin{equation}
\mathbf{\Psi}(t)\;=\;U(t) \mathbf{\Psi}(0) \;\;\; {\rm with} \;\;\; U(t)\;=\;\exp(t{H})\,,
\label{psuh}
\end{equation}
where the matrix $H$ is represented by two parts,
\begin{equation}
H\;=\;H_1\,+\,H_2\,,
\label{mahgen}
\end{equation}
with
\begin{eqnarray}
H_{1} &=&
\mathop{{\sum}'}_{i=1}^{n-2}
\beta_{i+1,i}\left(
\be^{\pht}_{i}\be^T_{i+1}-
\be^{\pht}_{i+1}\be^T_{i}
\right)\,,\label{h1}\\
H_{2} &=&
\mathop{{\sum}'}_{i=1}^{n-2}
\beta_{i+1,i+2}\left(
\be^{\pht}_{i+1}\be^T_{i+2}-
\be^{\pht}_{i+2}\be^T_{i+1}
\right)\,.
\label{h2}
\end{eqnarray}
Here, $\beta_{i,j}=1/(\delta\sqrt{\varepsilon_i \mu_j})$ and the prime
indicates that the sum is over odd integers only.
For $n$ odd we have
\begin{equation}
\dd{t}\Psi(1,t)=\beta_{2,1}
\Psi(2,t)\;\;\;\;\;{\rm and}\;\;\;\;\;
\dd{t}\Psi(n,t)=-\beta_{n-1,n}
\Psi(n-1,t)\,,
\end{equation}
such that the electric field vanishes at the boundaries
($Y_z(0,t)=Y_z(n+1,t)=0$), as required by the boundary conditions
Eqs.~(\ref{boundcond}).

The representation of $H$ as the sum of $H_{1}$ and $H_{2}$ divides the
lattice into odd and even numbered cells.
Most important, however, both $H_{1}$ and $H_{2}$ are skew-symmetric
block-diagonal matrices, containing one $1\times1$ matrix and $(n-1)/2$ real
$2\times2$ skew-symmetric matrices.
Therefore, according to the general theory outlined in Sec.~\ref{sec2},
this decomposition of $H$ is suitable to construct an orthogonal approximation
\begin{equation}
U_1(\tau)=e^{\tau H_{1}} e^{\tau H_{2}}
\label{U1}
\end{equation}
that is identical to the time-evolution operator $U(\tau)$ up to first order
in $\tau$.
As the matrix exponential of a block-diagonal matrix is equal to the
block-diagonal matrix of the matrix exponentials of the individual blocks,
the numerical calculation of $e^{\tau H_{1}}$ (or $e^{\tau H_{2}}$) reduces
to the calculation of $(n-1)/2$ matrix exponentials of $2\times2$ matrices.
The matrix exponential of a typical $2\times2$ matrix appearing in
$e^{\tau H_{1}}$ or $e^{\tau H_{2}}$ is simply given by
\begin{eqnarray}
\exp\left[\alpha
\left(\begin{array}{cc} \php0&1\\ -1&0\end{array}\right)\right]
\left(\begin{array}{c} \Psi(i,t) \\ \Psi(j,t) \end{array} \right)
\label{twobytwo} % \\
&=&\left(
\begin{array}{cc}
\php\cos \alpha & \php\sin\alpha
\\ -\sin \alpha & \php\cos\alpha
\end{array}\right)
\left(\begin{array}{c} \Psi(i,t) \\ \Psi(j,t) \end{array} \right)\,,
\label{rotitman}
\end{eqnarray}
and represents the rotation of two elements of the vector $\mathbf{\Psi}(t)$
leaving all the other elements unchanged.
This property of the time-evolution operator Eq.~(\ref{U1}) provides the
intrinsic possibility to parallelize the algorithms.
Furthermore, it is even possible to alter the ordering of the products in
the time-evolution operator $U_n(\tau)$ in order to construct an efficient
implementation for a particular system.
The plane rotations Eq.~(\ref{rotitman}) are performed by simply processing
an arbitrarily ordered list $S$ of pairs of EM field vector elements using
\begin{equation}
U_1(\tau)\;=\;\prod_{S}\! e^{\tau \beta_{i,j}
(\be^{\pht}_{i}\be^T_{j}-
\be^{\pht}_{j}\be^T_{i})}\,,
\label{U1S}
\end{equation}
instead of the odd-even decomposition Eq.~(\ref{U1}) for which
$S=\{(1,2), (3,4), \ldots, (n-2,n-1), (2,3), (4,5), \ldots, (n-1,n)\}$.

The implementation for 1D can be readily extended to 2D and 3D systems, as
has been illustrated in Ref.~\cite{Kole01}.
In 2D, the TDME (\ref{max}) separate again into two independent sets of
equations and the discretization of continuum space is done by simply
reusing the 1D lattice introduced above.
\begin{figure}[h]
\begin{center}
\includegraphics[width=8.0cm]{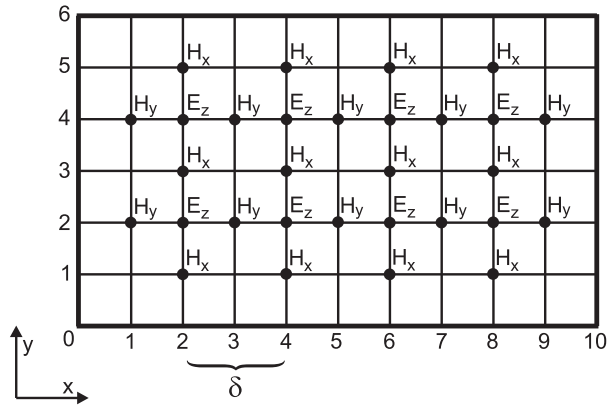}
\caption{Positions of the three TM-mode EM field components on the
2D grid for $n_x=9$ and $n_y=5$.
\label{fig:2DTM}}
\end{center}
\end{figure}
This is shown in Fig.~\ref{fig:2DTM} for the case of the 2D TM-modes.
The construction automatically takes care of the boundary conditions if
$n_x$ and $n_y$ are odd and yields a real skew-symmetric matrix $H$.
Correspondingly, in 3D the spatial coordinates are discretized by adopting
the standard Yee grid~\cite{Yee66}, which also automatically satisfies the
boundary conditions Eqs.~(\ref{boundcond}).
A unit cell of the Yee grid is shown in Fig.~\ref{fig:3D}.
\begin{figure}[h]
\begin{center}
\includegraphics[width=8.0cm]{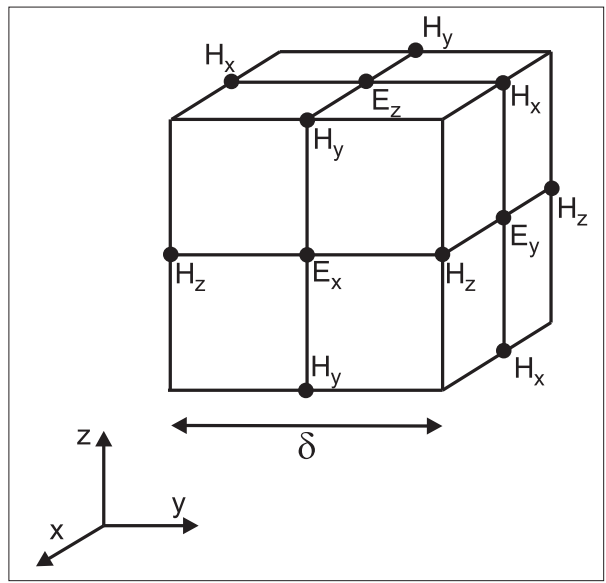}
\caption{Positions of the EM field components on the 3D Yee grid.
\label{fig:3D}}
\end{center}
\end{figure}
%

%
\iffalse
%
It is convenient to introduce a notation that specifies an unconditionally
stable algorithm by its temporal and spatial accuracy.
If the implementation of the algorithm involves a time integration based
on $U_n(\tau)$ and a spatial discretization scheme based on the $m$th-order
accurate approximation of the spatial derivatives, we will refer to this
algorithm as T$n$S$m$.
%
\fi
%

\begin{figure}[h]
\begin{center}
\includegraphics[width=8.0cm]{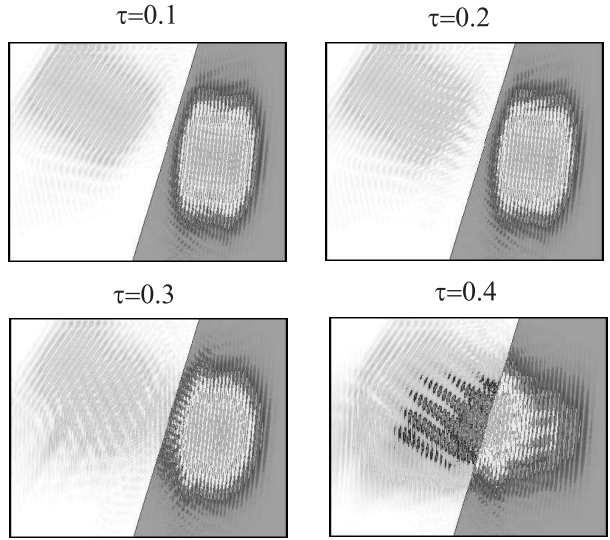}
\medskip
\caption{Energy density distributions at simulation time $t=12.8$
for various time steps $\tau$ obtained by the T4S2 algorithm for
a fixed mesh size $\delta=0.1$.
The wave packet has been created for parameters
$(\sigma_x,\sigma_y)=(2,1.73)$, $(x_0,y_0)=(5,7.5)$, and $k=8$
(see for details Eq.~(\ref{eqn:field_2d}) in Sec.~\ref{sec4_3})
and impinges on the dielectric structure from the left.
\label{fig:courant2d}}
\end{center}
\end{figure}

In general, the time step $\tau$ and the distance $\delta$ between
next-nearest neighbor grid points are related due to the error that is
introduced when the exact time-evolution operator $U(\tau)$ is replaced by
$U_n(\tau)$.
We have~\cite{Suzuki8591,DeRaedt83,DeRaedt87}:
\begin{equation}
\|U(\tau)-U_n(\tau)\|\,\leq\,
\gamma(d)\left(\frac{\alpha(n)\,\tau}{\delta}\right)^{n+1}.
\label{reserror}
\end{equation}
Here, $\gamma(d)$ depends on the particular spatial discretization scheme used
and $\alpha(n)$ represents the largest positive constant that appears as a
prefactor in the exponential of the approximation $U_n(\tau)$.
We find $\alpha(2)=1/2$ from Eq.~(\ref{secordapp}) and inspection of
Eq.~(\ref{fouordapp}) yields $\alpha(4)=(1/2)(4a-1)\approx 0.33$.
It follows that for a required spatial resolution, which determines the
smallness of $\delta$, the time step has to be chosen such that
\begin{equation}
\tau\;\leq\;\tau^\ast\;\equiv\;\frac{\delta}{\alpha(n)}
\label{ourcour}
\end{equation}
in order to keep the error Eq.~(\ref{reserror}) small.
As an example we consider a wave packet in a 2D cavity that is simulated
by a T4S2 algorithm.
For numerical purposes we use dimensionless variables throughout this
paper, where the unit of length is denoted by $\lambda$ and the vacuum
light velocity $c$ is taken as the unit of velocity, while the permittivity
$\varepsilon$ and permeability $\mu$ are measured in units of their
corresponding values in vacuum, respectively, $\varepsilon_0$ and
$\mu_0$.
The cavity with $\varepsilon=1$ and $\mu=1$ is of size $19\times 15$ and
contains a dielectric medium with $\varepsilon=2.25$ and $\mu=1$ that
has an inclined boundary.
We plot in Fig.~\ref{fig:courant2d} the results of simulations in which
the wave packet scatters on the dielectric medium.
In the four pictures we show the EM energy density distributions that
are obtained after simulation time $t=12.8$ for a fixed mesh size
$\delta=0.1$ and for four different time steps $\tau$.
It follows from Eq.~(\ref{ourcour}) that the upper limit for the time step
is given by $\tau^\ast=0.3$ in this case.
For $\tau=0.4$ the EM energy density distribution is, in fact, seen to change
dramatically such that the results become meaningless.
It should be noted that the limitation Eq.~(\ref{ourcour}) on the time
step is different from the Courant number which relates the time step
$\tau$ to the stability of finite-difference time-domain (FDTD)
algorithms~\cite{Taflove} that are based on the Yee algorithm~\cite{Yee66}.
The algorithms presented in this paper are unconditionally stable by
construction for any time step $\tau$ and produce reasonable numerical
results up to $\tau=\tau^\ast$, a time step at which the Yee-based FDTD
algorithms may have become unstable.

\begin{figure}[h]
\begin{center}
\includegraphics[width=8.0cm]{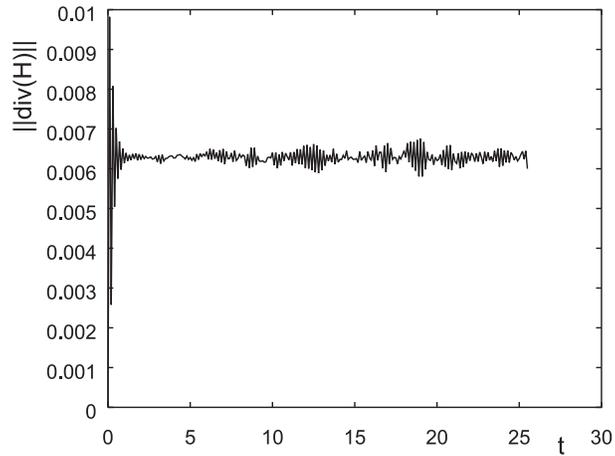}
\medskip
\caption{The norm of the divergence of the magnetic field in a 3D empty cavity
($\varepsilon=1$ and $\mu=1$) of size $1.5 \times 1.5 \times 1.5$ as a function
of time $t$.
The computation is performed with the T2S2 algorithm keeping the mesh size
$\delta=0.1$ fixed.
\label{fig:div3d_time}}
\end{center}
\end{figure}

We conclude this section by noting that our algorithms conserve the divergence
of the EM fields only in 1D systems but not in 2D and 3D systems.
Although the initial state $\mathbf{\Psi}(t=0)$ can be chosen
such that the EM fields satify Eqs.~(\ref{maxdiv}), the time-integration of
the TDME by an algorithm based on the approximation $U_n(\tau)$ yields EM
fields whose divergence quickly acquires a finite value and then remains
constant in time.
This is shown in Fig.~\ref{fig:div3d_time} where we plot the computed norm
of the magnetic field divergence in a 3D system as a function of time.
The 3D system is an empty cavity ($\varepsilon=1$
and $\mu=1$) of size $1.5 \times 1.5 \times 1.5$ and we use the
T2S2 algorithm.
\begin{figure}[h]
\begin{center}
\includegraphics[width=8.0cm]{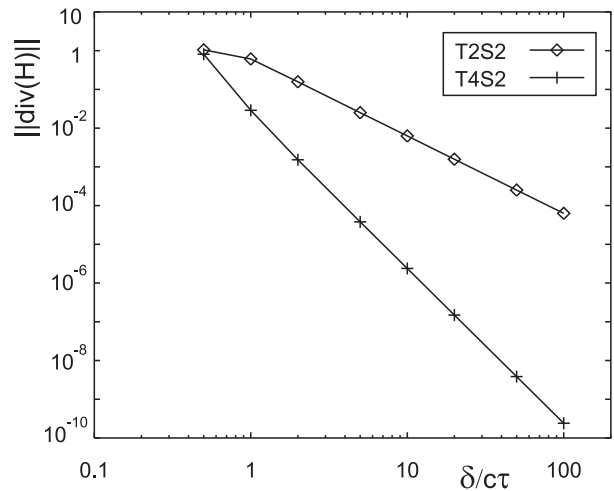}
\medskip
\caption{The norm of the divergence of the magnetic field in a 3D empty cavity
($\varepsilon=1$ and $\mu=1$) of size $1.5 \times 1.5 \times 1.5$ as a function
of $\delta/c\tau$.
The computation is performed with the algorithms T2S2 and T4S2.
\label{fig:divergence3d}}
\end{center}
\end{figure}
Though the divergence of the EM fields is not conserved in 2D and 3D systems,
this error can be reduced by using smaller time steps or algorithms with
higher-order time accuracy.
This can be seen in Fig.~\ref{fig:divergence3d}, where we compare the algorithms
T2S2 and T4S2 as a function of the time step $\tau$ to show that the error in the
EM field divergence vanishes for the T$n$S$2$ algorithm proportional to
$(\tau c/\delta)^n$.

%
%%%%%%%%%%%%%%%%%%%%%%%%%%%%%%%%%%%%%%%%%%%%%%%%%%%%%%%%%%%%%%%%%%%%%%%%%%%%
%
\section{Variable grid implementation}\label{sec4}
%
%%%%%%%%%%%%%%%%%%%%%%%%%%%%%%%%%%%%%%%%%%%%%%%%%%%%%%%%%%%%%%%%%%%%%%%%%%%%
%

The poor man's implementation does not provide an optimal discretization
scheme for physical systems of unregular geometrical shapes or with strongly
varying permeability and/or permittivity.
In a practical implementation of such systems the grid has to be variable
with a small mesh size in one region of the system and a large mesh size
in another region of the system.
In this section we show how to implement a variable grid in such a way that
the algorithms to solve the TDME remain unconditionally stable by
construction.

\begin{figure}[h]
\begin{center}
\includegraphics[width=8.0cm]{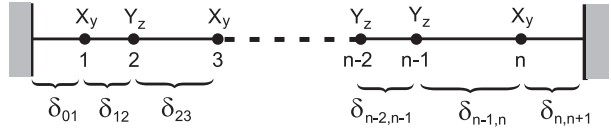}
\medskip
\caption{Positions of the two TM-mode EM field components on the
1D variable grid.
\label{fig:varmesh1D}}
\end{center}
\end{figure}

For the sake of simplicity we consider a 1D system that is discretized
using a variable grid as shown in Fig.~\ref{fig:varmesh1D}.
In a straightforward implemention of the variable grid we would replace
the constant next-nearest neighbor distance $\delta$ in
Eqs.~(\ref{eqn:discrTMX}) and (\ref{eqn:discrTMY}) of the poor man's
implementation by the corresponding variable distance.
It is convenient to write this substitution in the form
\begin{equation}
\delta\;\rightarrow\;\Delta_{i,i+1}
\left[1\,+\,
\frac{\delta_{i-1,i}-\delta_{i+1,i+2}}{2\Delta_{i,i+1}}\right]\,,
\label{replaceit}
\end{equation}
where $\delta_{i,j}$ is the distance between grid points $i$ and $j$
(see Fig.~\ref{fig:varmesh1D}) and
\begin{equation}
\Delta_{i,i+1}\;\equiv\;
\frac{1}{2}(\delta_{i-1,i}\,+\,2\delta_{i,i+1}\,+\,\delta_{i+1,i+2})
\end{equation}
is the averaged next-nearest neighbor distance.
It can be easily checked that an implementation of the variable grid that
relies on the replacement Eq.~(\ref{replaceit}) would destroy the
skew-symmetry property of the corresponding matrix $H$ (see
Eq.~(\ref{mahgen})).
This is unphysical: The original form of the Maxwell equations do have this
property.
However, a variable grid implementation that does preserve the underlying
symmetry of Maxwell's equations can be constructed for a sufficiently smooth,
variable grid.
In this case, the second term in the brackets of Eq.~(\ref{replaceit}) may be
neglected and the replacement
\begin{equation}
\delta\;\rightarrow\;\Delta_{i,i+1}\;=\;\Delta_{i+1,i}
\label{replaceapprox}
\end{equation}
may yield a resonable approximation of Eqs.~(\ref{eqn:discrTMX}) and
(\ref{eqn:discrTMY}) for the variable grid implementation:
\begin{eqnarray}
\dd{t}X_y(i,t) & = & \frac{1}{\sqrt{\mu_i}} \left(
\frac{Y_z(i+1,t)}{\Delta_{i,i+1}\sqrt{\varepsilon_{i+1}}}-
\frac{Y_z(i-1,t)}{\Delta_{i,i-1}\sqrt{\varepsilon_{i-1}}}\right)\,,
\label{newdiscrTMX}
\\ \dd{t}Y_z(i+1,t) & = & \frac{1}{\sqrt{\varepsilon_{i+1}}} \left(
\frac{X_y(i+2,t)}{\Delta_{i+1,i+2}\sqrt{\mu_{i+2}}}-
\frac{X_y(i,t)}{\Delta_{i+1,i}\sqrt{\mu_{i}}}\right)\,.
\label{newdiscrTMY}
\end{eqnarray}
The corresponding matrix $H$ is seen to be skew-symmetric,
\begin{equation}
H= \mathop{{\sum}'}_{i=1}^{n} \left[
\frac{\be^{\pht}_{i}\be^T_{i+1}-\be^{\pht}_{i+1}\be^T_{i}}
{\Delta_{i,i+1}\sqrt{\varepsilon_{i+1}\mu_i}} +
\frac{\be^{\pht}_{i+1}\be^T_{i+2}-\be^{\pht}_{i+2}\be^T_{i+1}}
{\Delta_{i+1,i+2}\sqrt{\varepsilon_{i+1}
\mu_{i+2}}} \right]\,, \label{eqn:oper_1D_new}
\end{equation}
and may again be separated into an odd and even part of which the exponents
can be easily calculated following the same steps as given above in the poor
man's implementation.

It is obvious that this variable grid implementation can, in principle, be
applied in any spatial dimension $d$.
However, it is in general not possible to predict how to choose a grid that
yields the best approximation to the true spectrum of eigenmodes of any
non-trivial $d$-dimensional system.
We therefore studied the criteria for the choice of suitable variable grids
in particular systems numerically and present the results for a 1D and a 2D
system in the remainder of this section.

\begin{figure}[h]
\begin{center}
\includegraphics[width=8.0cm]{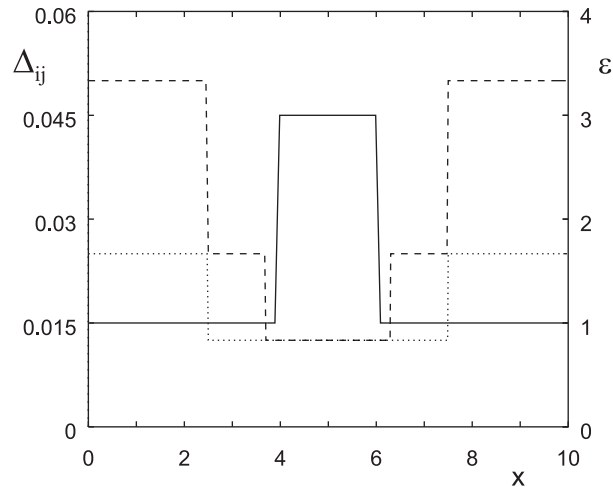}
\medskip
\caption{
The 1D cavity with the dielectric structure (solid line) and the two
implemented variable grids:
$\Delta_{i,i+1}=\{0.1\leftrightarrow 0.05\leftrightarrow 0.025\}$
(dashed line) and $\Delta_{i,i+1}=\{0.05\leftrightarrow 0.025\}$
(dotted line). \label{fig:vargridsys}}
\end{center}
\end{figure}

The 1D system under consideration consists of a cavity of length $L=10$ with
a constant permeability $\mu=1$ and a varying permittivity $\varepsilon$.
The permittivity deviates from its vacuum value ($\varepsilon=1$) due to the
presence of a dielectric medium with $\varepsilon=3$ that is located in the
middle of the cavity and extends over a length $2$, as shown in
Fig.~\ref{fig:vargridsys}.
As a reference system we use a poor man's implementation with constant
next-nearest neighbor distance $\delta=0.025$ and calculate the eigenmodes
$\omega_n$ of the corresponding matrix $H$.
For two different variable grids we calculate the corresponding eigenmodes
$\tilde{\omega}_n$ and the deviation
$\Gamma(\omega_n,\tilde{\omega}_n)=1-\tilde{\omega}_n/\omega_n$
relative to the eigenmodes of the reference system.
The two variable grids have in common that the dielectric medium and the
transitions between $\varepsilon=1$ and $\varepsilon=3$ at both its sides
is embedded in a grid of constant next-nearest neighbor distance which equals
that of the reference system ($\Delta_{i,i+1}=\delta=0.025$).
Furthermore, at the left end and at the right end of the cavity the
next-nearest neighbor distance is constant over a length $2.5$ and equals,
respectively, $\Delta_{i,i+1}=0.1$ and $\Delta_{i,i+1}=0.05$ in the two
variable grids.
The transitions in the variable grids between regions of constant next-nearest
neighbor distance involve abrupt steps between
\begin{equation}
\Delta_{i,i+1}\,=\,0.1\; \leftrightarrow \;\Delta_{i,i+1}\,=\,0.05\;
\leftrightarrow \;\Delta_{i,i+1}\,=\,0.025\,,
\end{equation}
where we kept the intermediate distance $\Delta_{i,i+1}=0.05$ over eight grid
points,
and between
\begin{equation}
\Delta_{i,i+1}\,=\,0.05 \;\leftrightarrow\; \Delta_{i,i+1}\,=\,0.025\,,
\end{equation}
respectively.

\begin{figure}[h]
\begin{center}
\includegraphics[width=8.0cm]{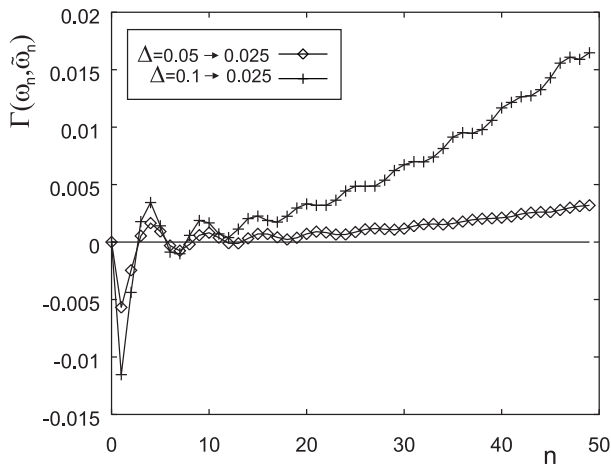}
\medskip
\caption{
Relative deviation $\Gamma(\omega_n,\tilde{\omega}_n)$ for two variable grids.
\label{fig:masym}}
\end{center}
\end{figure}

\begin{figure}[h]
\begin{center}
\includegraphics[width=8.0cm]{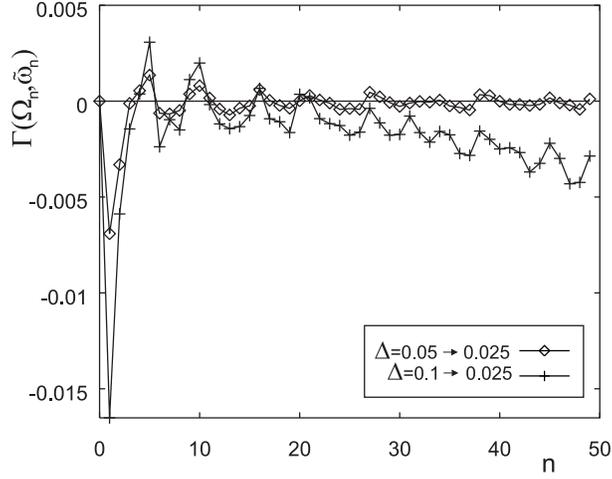}
\medskip
\caption{
Relative deviation $\Gamma(\Omega_n,\tilde{\omega}_n)$ for two variable grids.
\label{fig:manosym}}
\end{center}
\end{figure}

In Fig.~\ref{fig:masym} we plot $\Gamma(\omega_n,\tilde{\omega}_n)$ for the
first 50 eigenmodes of both variable grids.
The relative deviation is seen to increase with the number of the frequency
mode.
As high mode numbers represent high frequencies this observation simply
reflects the general fact that the accuracy of the eigenmodes depends on
the smallness of the mesh size (numerical dispersion).
Clearly, this also explains why the relative deviation
$\Gamma(\omega_n,\tilde{\omega}_n)$ increases upto $2\%$ for the variable
grid with $\Delta_{i,i+1}=\{0.1\leftrightarrow 0.05\leftrightarrow 0.025\}$,
while for the variable grid with $\Delta_{i,i+1}=\{0.05\leftrightarrow 0.025\}$
this deviation remains well below $0.5\%$.
For the first few frequency modes, however, we observe an increase in
$\Gamma(\omega_n,\tilde{\omega}_n)$.
This behavior can be related to the error that is introduced in the variable
grid implementation by applying the approximation Eq.~(\ref{replaceapprox})
instead of the exact replacement Eq.~(\ref{replaceit}).
To check this statement we plot in Fig.~\ref{fig:manosym} the deviation
$\Gamma(\Omega_n,\tilde{\omega}_n)$ for the first 50 eigenmodes of the two
variable grids relative to the eigenmodes $\Omega_n$ that belong to the
variable grids of the exact implementation Eq.~(\ref{replaceit}).
We see that the increase of the relative deviation for the first few
eigenmodes is, in fact, related to the error which is made by replacing the
exact substitution Eq.~(\ref{replaceit}) with the skew-symmetry conserving
approximation Eq.~(\ref{replaceapprox}).
This approximation leads to oscillations of
$\Gamma(\Omega_n,\tilde{\omega}_n)$
(and also $\Gamma(\omega,\tilde{\omega}_n)$) that vanish with increasing
frequency mode number.
From extended numerical studies (results not shown) we find that these
variations depend on several factors, such as the size in the difference
between the largest and smallest distance $\Delta_{i,i+1}$ of the variable
grid implementation and on how abrupt $\Delta_{i,i+1}$ changes with $i$.
In practice, it will be necessary to check the robustness of numerical
results obtained by a variable grid implementation against small changes in
its parameters.
Although this may sound as a serious disadvantage, the next example of a 2D
system shows that for realistic applications it may be by far more efficient
to perform several simulation runs with a variable grid implementation than
to use the poor man's implementation.

\begin{figure}[h]
\begin{center}
\includegraphics[width=4.0cm]{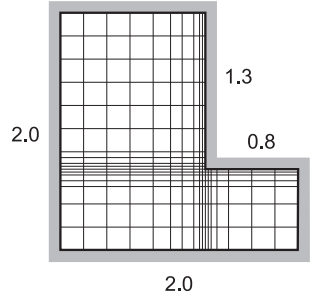}
\medskip
\caption{The L-shaped 2D cavity with a variable grid (schematically).
\label{fig:cav2d}}
\end{center}
\end{figure}

The 2D system we consider is given by the L-shaped cavity depicted in
Fig.~\ref{fig:cav2d}.
In order to satisfy the conditions Eq.~(\ref{boundcond}) at the boundaries, the
EM fields change very strongly close to the sharp edge of the cavity.
Large spatial changes of the EM fields require a small mesh size.
However, for the overwhelming part of the cavity a small mesh size would cause
a waste of resources (computer memory and CPU time).
Therefore, this system can be more efficiently simulated by a variable grid
implementation with an increasing number of grid points near the edge.
This is done by a uniform increase of the number of grid points along both the
$x$- and the $y$-direction as is schematically drawn in Fig.~\ref{fig:cav2d}.
Furthermore, instead of using the odd-even decomposition of the time-evolution
operator (corresponding to Eq.~(\ref{U1}) for the 1D system) on a square grid
that would contain grid points outside the L-shaped cavity,
we perform the plane rotations by processing a list $S$ of pairs of the EM field
vector elements at the grid points that actually belong to the L-shaped cavity
(corresponding to Eq.~(\ref{U1S}) for the 1D system).

\begin{table}
\begin{tabular}{ccccc}
mode $n$ & \multicolumn{2}{c}{T2S2} & \multicolumn{2}{c}{GdfidL} \\
    & constant & variable & constant & variable \\
    & grid $\omega_n$ & grid $\omega_n$ & grid $\omega_n$ & grid $\omega_n$ \\
    \hline
1   & 2.9989 & 2.9913 & 2.9999 & 2.9992 \\
2   & 3.9807 & 3.9500 & 3.9740 & 3.9720 \\
3   & 4.9164 & 4.8857 & 4.9156 & 4.9102 \\
4   & 5.4150 & 5.3843 & 5.4077 & 5.4004 \\
5   & 5.5837 & 5.5453 & 5.5791 & 5.5710 \\
6   & 6.0592 & 6.0209 & 6.0580 & 6.0494 \\
7   & 6.7649 & 6.7265 & 6.7511 & 6.7377 \\
8   & 6.8876 & 6.8492 & 6.8797 & 6.8674
\end{tabular}
\caption{The eight lowest TM eigenmodes of the L-shaped cavity
(see Fig.~\ref{fig:cav2d}).}
\label{tbl:tbl1}
\end{table}
In Table~\ref{tbl:tbl1} we present the results of a numerical simulation
for the eight lowest TM eigenmodes in the cavity.
We used the T2S2 algorithm imposing a poor man's implementation with
$\delta=0.003125$ and a variable grid implementation with a mesh size
ranging from $\Delta=0.05$ to $\Delta=0.003125$.
Very similar to the procedure described above for the 1D system, the mesh size
is decreased by a factor $0.5$ and then kept constant for several grid points
to smoothen this transition before the mesh size is decreased further.
Our results are in good agreement with those obtained by the program package
GdfidL~\cite{GdfidL} for the same 2D system (see Table~\ref{tbl:tbl1}).
In Table~\ref{tbl:tbl2} we show the location of the arbitrarily chosen
third-lowest eigenmode $\omega_3$ for several constant and variable grid
implementations of the T2S2 algorithm.
In all simulations we set $\delta/c\tau=10$, where in the case of a variable
grid $\delta$ is replaced by the smallest mesh size.
The relative error $\Gamma$ of the frequency $\omega_3$ is measured with
respect to the frequency $\omega_3=4.916$ of the system with constant mesh
size $\delta=0.003125$.
\begin{table}
\begin{tabular}{lcc}
constant grid $\delta$   & $\omega_3$ & $\Gamma$ (in \%) \\ \hline
0.1         & 4.571 & 7.5  \\
0.05        & 4.740 & 3.7  \\
0.025       & 4.832 & 1.7  \\
0.0125      & 4.878 & 0.78 \\
0.00625     & 4.901 & 0.31 \\
0.003125    & 4.916 & 0    \\ \hline
variable grid $\Delta$   &       &      \\ \hline
$0.1 \rightarrow 0.05       $ & 4.717 & 4.2  \\
$0.1 \rightarrow 0.025      $ & 4.801 & 2.4  \\
$0.1 \rightarrow 0.0125     $ & 4.840 & 1.6  \\
$0.1 \rightarrow 0.00625    $ & 4.878 & 0.78 \\
$0.05 \rightarrow 0.003125  $ & 4.886 & 0.61 \\
\end{tabular}
\caption{Error in third-lowest eigenmode of the L-shaped cavity
(see Fig.~\ref{fig:cav2d}).}
\label{tbl:tbl2}
\end{table}
The numerical results obtained within the variable grid implementation are
in excellent agreement with the results of the poor man's implementation
and the program package GdfidL.
The T2S2 algorithm with the poor man's implementation and $\delta=0.003125$
consumes about 150 times more CPU time and 10 times more computer memory than
the T2S2 algorithm with variable grid implementation and
$\Delta=\{0.05\rightarrow 0.003125\}$.
Clearly, these numbers justify additional simulation runs that are required
to check the robustness of numerical results against small changes in the
parameters of a variable grid implementation.

%
%%%%%%%%%%%%%%%%%%%%%%%%%%%%%%%%%%%%%%%%%%%%%%%%%%%%%%%%%%%%%%%%%%%%%%%%%%%%
%
\section{Improved spatial discretization implementation}\label{sec5}
%
%%%%%%%%%%%%%%%%%%%%%%%%%%%%%%%%%%%%%%%%%%%%%%%%%%%%%%%%%%%%%%%%%%%%%%%%%%%%
%

Both conditional FDTD algorithms and the unconditionally stable algorithms
T$n$S$m$ suffer from numerical dispersion due to the discretization of continuum
space on a grid with a finite mesh size~\cite{Taflove}.
Methods to reduce numerical dispersion are taking a grid with a smaller mesh
size or employing more accurate finite-difference approximations to the
spatial derivatives.
The former obviously can be also used in the poor man's implementation of
unconditionally stable algorithms, however, for several reasons it may be
more desirable to implement higher-order accurate approximations of the spatial
derivatives.
For example, if one is interested in global features of the distribution of
a system's eigenmodes, i.e. if we want to determine {\it all} eigenvalues,
a higher-order accurate spatial derivative implementation would be strongly
preferred.
The computation of a system's eigenmode spectrum is performed by calculating
the Fourier transform of the inner product
$F(t)=\langle\mathbf{\Psi}(0)|\mathbf{\Psi}(t)\rangle$~\cite{Kole01,Alben75,Hams00}.
Using independent random numbers to initialize the elements of
$\mathbf{\Psi}(0)$, the full eigenmode spectrum is obtained by averaging this
Fourier transform.
Taking just a smaller mesh size for the grid in the poor man's implementation
does not only reduce the numerical dispersion but also gives rise to more
eigenmodes with high frequencies.
In order to obtain the eigenmode spetrum with the same spectral resolution,
the sampling of $F(t)$ would have to be done over smaller time intervals
involving the computation of more data points.
It is thus desirable to implement, instead, higher-order accurate approximations
of the spatial derivatives that make a moderate use of computer resources in
terms of CPU time and computer memory possible.

The procedure for the construction of higher-order approximations to spatial
derivatives is standard~\cite{num_anal}.
In the present case, we apply this procedure keeping in mind that Maxwell's
equations~(\ref{eqn:mtxeqn}) are skew-symmetry and that the electric and
magnetic field components are defined at particular grid points.
The grid of a $d$-dimensional system with a constant mesh size of distance
$\delta/2$ between neighboring grid points is shown in
Figs.~(\ref{fig:fig1})-(\ref{fig:3D}).
Without loss of generality we consider a 1D system, where
$\Psi(i,t)=\Psi(i\delta/2,t)$ is the $i$th component of the EM field vector
and denotes an electric field compoenent for $i$ even and a magnetic field
component for $i$ odd (see Sec.~\ref{sec3} for details).
Applying the second-order accurate central-difference scheme the spatial
derivative of the EM field component $\Psi(i,t)$ is given by
\begin{equation}
\dd{x}\Psi(i,t)\;=\;
\frac{\Psi(i+1,t)-\Psi(i-1,t)}{\delta}-\frac{\delta^2}{6}\Psi^{(3)}(i,t)
+{\mathcal O}(\delta^4),
\label{eqn:4th_1}
\end{equation}
where $\Psi^{(3)}(i,t)\equiv \partial^3 \Psi(i,t)/\partial x^3$.
Similarly, using the third-nearest neighbor EM field points at distance
$3\delta/2$, we have
\begin{equation}
\dd{x}\Psi(i,t)\;=\;
\frac{\Psi(i+3,t)-\Psi(i-3,t)}{3\delta}-\frac{9\delta^2}{6}\Psi^{(3)}(i,t)
+{\mathcal O}(\delta^4).
\label{eqn:4th_2}
\end{equation}
A fourth-order accurate approximation of the spatial derivative
$\partial\Psi(i,t)/\partial x$ is now constructed in terms of a linear
combination of Eqs.~(\ref{eqn:4th_1}) and (\ref{eqn:4th_2}) which is chosen
such that the terms proportional to $\Psi^{(3)}(i,t)$ vanish.
We obtain:
\begin{equation}
\dd{x}\Psi(i,t)\;=\;
\frac{9}{8}\left(\frac{\Psi(i+1,t)-\Psi(i-1,t)}{\delta}\right)-
\frac{1}{8}\left(\frac{\Psi(i+3,t)-\Psi(i-3,t)}{3\delta}\right)
+{\mathcal O}(\delta^4).
\label{reshighord}
\end{equation}
In practice, it is straightforward to implement the improved spatial
discretization, since we can use the implementation of the central-difference
scheme for the two terms separately and then combine the results according
to Eq.~(\ref{reshighord}).
The corresponding matrix $H$ of the 1D system (see Eq.~(\ref{mahgen})) changes
from tridiagonal to five-diagonal, but most importantly it
preserves its property of being skew-symmetric.
It should be noted, however, that the fourth-order accurate spatial
derivative introduces errors at the boundaries since the calculation of
$\partial\Psi(i,t)/\partial x$ for
$i=1$,$2$,$n-1$, and $n$ refer, respectively, to grid points
$i=-2$, $-1$, $n+1$, and $n+2$ that lie outside the cavity and are
implicitly assumed to be zero.

It is obvious that the fourth-order accurate approximation of the spatial
derivatives can be similarly applied in systems of any spatial dimension $d$.
In the remainder of this section we study the numerical dispersion and the
temporal and spatial accuracy of the algorithms for various 1D and 2D systems.

%%%%%%%%%%%%%%%%%%%%%%%%%%%%%%%%%%%%%%%%%%%%%%%%%%%%%%%%%%%%%%%%%%%%%%
%
\subsection{Numerical Dispersion}\label{secnumdisp}
%
%%%%%%%%%%%%%%%%%%%%%%%%%%%%%%%%%%%%%%%%%%%%%%%%%%%%%%%%%%%%%%%%%%%%%%

\begin{figure}[h]
\begin{center}
\includegraphics[width=8.0cm]{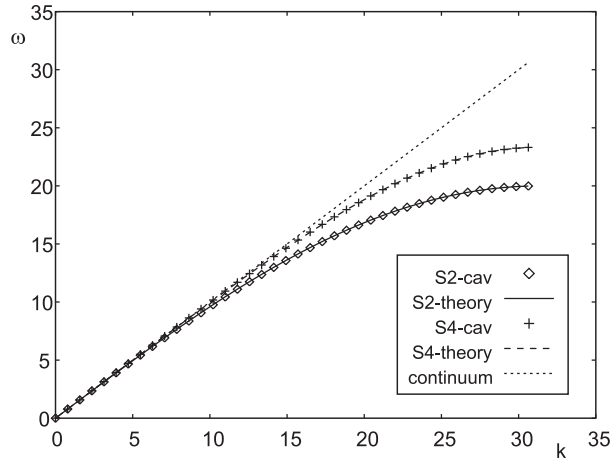}
\medskip
\caption{Numerical and analytical dispersion relations for the 1D cavity
of length $L=4$ as obtained from calculations with $m$th-order accurate
approximations of the spatial derivatives ($m=2,4$).
In both simulations we kept $\delta=0.1$ and $\tau=0.01/c$ fixed.
\label{fig:dispersion}}
\end{center}
\end{figure}

We illustrate the difference in the numerical dispersion between
the poor man's implementation and the improved spatial discretization
implementation by a comparison of the eigenmode spectra of a 1D empty cavity
($\varepsilon=1$ and $\mu=1$) of length $L$.
In 1D, the continuum wave equation for the EM fields~\cite{BornWolf},
\begin{equation}
\left[\frac{1}{c^2}\frac{\partial^2}{\partial
t^2}\,-\,\frac{\partial^2}{\partial x^2}\right]\Psi(x,t)\;=\;0\,,
\label{waffwaff}
\end{equation}
is solved by the ansatz $\Psi(x,t)\propto\cos(\omega t-kx+\phi)$
(with a phase $\phi$ to distinguish electrical and magnetic field
components)
yielding the linear dispersion relation between frequency $\omega$
and wave number $k$: $\omega=c|k|$.
Focusing on the effect of the spatial derivatives on the numerical
dispersion, we assume perfect time integration of the algorithms and
impose periodic boundary conditions on the EM field components:
$\Psi_p(i,t)\propto\cos(\omega_p t- k_p \delta/2 +\phi)$
with wave number $k_p=2\pi p/L$ and $-L/(2\delta)<p\leq L/(2\delta)$.
Applying the second-order accurate spatial derivative we obtain
\begin{equation}
\frac{\partial^2}{\partial x^2}\Psi_p(i,t)=\frac{1}{\delta^2}
\left[\Psi_p(i+2,t)-2\Psi_p(i,t)+\Psi_p(i-2,t)\right]\,+\,
{\mathcal O}(\delta^2)\,,
\label{eqn:2disper}
\end{equation}
while for the fourth-order accurate spatial derivative we find
\begin{eqnarray}
\frac{\partial^2}{\partial x^2}\Psi_p(i,t)&=&
\left(\frac{9}{8\delta}\right)^2
\left[\Psi_p(i+2,t)-2\Psi_p(i,t)+\Psi_p(i-2,t)\right]+\nonumber\\
&+&\left(\frac{1}{24\delta}\right)^2
\left[\Psi_p(i+6,t)-2\Psi_p(i,t)+\Psi_p(i-6,t)\right]+\nonumber\\
&+&\left(\frac{9}{96\delta^2}\right)
\left[\Psi_p(i+2,t)+\Psi_p(i-2,t)-\Psi_p(i+4,t)-\Psi_p(i-4,t)\right]\,
+\,{\mathcal O}(\delta^4)\,.
\label{eqn:4disper}
\end{eqnarray}
For $m=2$ the analytical solution of the eigenmode spectrum for the
$m$th-order accurate spatial derivative is given by
\begin{equation}
{\omega_p}^2\;=\;2
\left(\frac{c}{\delta}\right)^2\left[1-\cos(k_p\delta)\right]\,,
\label{anasolum2}
\end{equation}
while for $m=4$ we find
\begin{equation}
{\omega_p}^2\;=\;
\left(\frac{c}{\delta}\right)^2
\sum_{l=0}^{3}C_l\cos(lk_p\delta)
\label{anasolum4}
\end{equation}
with coefficients $C_0=365/144$, $C_1=-87/32$,
$C_2=3/16$, and $C_3=-1/288$.
We show in Fig.~\ref{fig:dispersion} that the dispersion relations, which
we obtained numerically by the $m$th-order accurate spatial derivative
implementation for a 1D cavity of length $L=4$, are in excellent agreement
with the corresponding analytical solutions Eqs.~(\ref{anasolum2}) and
(\ref{anasolum4}).
It is clearly visible that the dispersion relation computed by the poor man's
implementation (T2S2 algorithm) suffers from numerical dispersion already at
frequencies above $\omega=10$, whereas for a grid with the same mesh size the
fourth-order accurate spatial derivative implementation (T2S4 algorithm) works
well up to $\omega=15$.

%
%%%%%%%%%%%%%%%%%%%%%%%%%%%%%%%%%%%%%%%%%%%%%%%%%%%%%%%%%%%%%%%%%
%
\subsection{Temporal and Spatial Accuracy}
\label{sec4_3}
%
%%%%%%%%%%%%%%%%%%%%%%%%%%%%%%%%%%%%%%%%%%%%%%%%%%%%%%%%%%%%%%%%%
%

To perform a systematic study of the accuracy of the algorithms as a function
of the time step $\tau$ and the mesh size $\delta$, we compute the difference
between the normalized exact, $\mathbf{\Psi}(t)$, and the approximate,
$\mathbf{\Psi}_{n,m}(t)$, EM field vector as obtained by the T$n$S$m$
algorithm:
\begin{equation}
\Delta\mathbf{\Psi}_{n,m}(t)\;\equiv\;
\|\mathbf{\Psi}(t)\,-\,\mathbf{\Psi}_{n,m}(t)\|\,.
\label{errortemspa}
\end{equation}

We first consider the propagation of a Gaussian wave packet in a 1D empty
cavity ($\varepsilon=1$ and $\mu=1$) of length $L=30$.
At $t=0$ the Gaussian wave packet
\begin{equation}
E_z(x,t)\;=\;\exp\left[-(x-x_0-ct)^2/\sigma^2\right] \label{eqn:movgau}
\end{equation}
with standard deviation $\sigma=2$ is located at $x_0=8$.
For $t>0$ the wave packet propagates with velocity $c$ in the $x$-direction
until it hits the right boundary of the cavity, becomes reflected, and
propagates in the opposite direction.
To derive an analytical expression of the exact EM field vector
$\mathbf{\Psi}(t)$, we expand $E_z(x,t)$ in the TM-modes
\begin{eqnarray}
E_z(x,t) & = & -\sum_{n=1}^{\infty} a_n \sin(n\pi x/L)\sin(n\pi (x_0+ct)/L),\\
H_y(x,t) & = & \frac{a_0}{2} +
               \sum_{n=1}^{\infty} a_n \cos(n\pi x/L)\cos(n\pi (x_0+ct)/L),
\label{eqn:fields_1d}
\end{eqnarray}
with coefficients
\begin{equation}
 a_n = \frac{2\sigma\sqrt{\pi}}{L}\exp\left( -\frac{\sigma^2}{4}
         \left(\frac{n\pi}{L}\right)^2\right),
\end{equation}
which ensure that the wave packet satisfies the boundary conditions
Eq.~(\ref{boundcond}).
Using Poisson's summation formula we find the following expressions
for the EM field components
\begin{eqnarray}
E_z(x,t) & = &
            \sum_{n=-\infty}^{\infty}\left[
\exp\left(-(2nL+x+x_0-ct)^2/\sigma^2\right)-
\exp\left(-(2nL+x-x_0+ct)^2/\sigma^2\right) \right]\,,
\\H_y(x,t) & = &  \sum_{n=-\infty}^{\infty}\left[
\exp\left(-(2nL+x+x_0-ct)^2/\sigma^2\right)+
\exp\left(-(2nL+x-x_0+ct)^2/\sigma^2\right) \right]\,,
\label{eqn:poisson_1d}
\end{eqnarray}
from which the exact EM field vector $\mathbf{\Psi}(t)$ is constructed
according to Eq.~(\ref{eqn:eqnindx}) on the 1D grid (see Fig.~\ref{fig:fig1}).

\begin{figure}[h]
\begin{center}
\includegraphics[width=8.0cm]{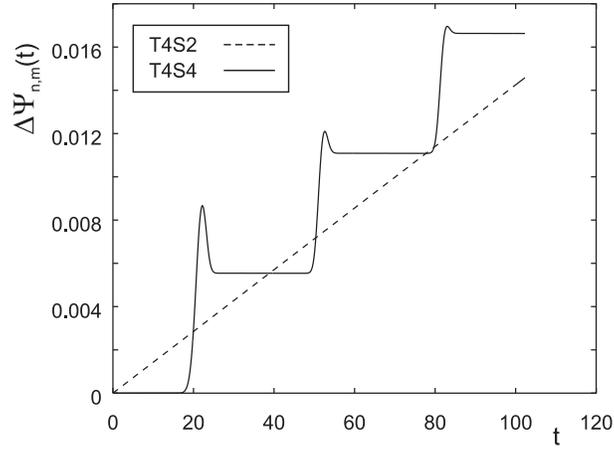}
\medskip
\caption{The error $\Delta\mathbf{\Psi}_{n,m}(t)$ as a function of the
simulation time $t$ for fixed values of the mesh size $\delta=0.1$ and the
time step $\tau=0.01/c$.
Results are shown for the T4S2 and T4S4 algorithm.
\label{fig:rms_time}}
\end{center}
\end{figure}

\begin{figure}[h]
\begin{center}
\includegraphics[width=8.0cm]{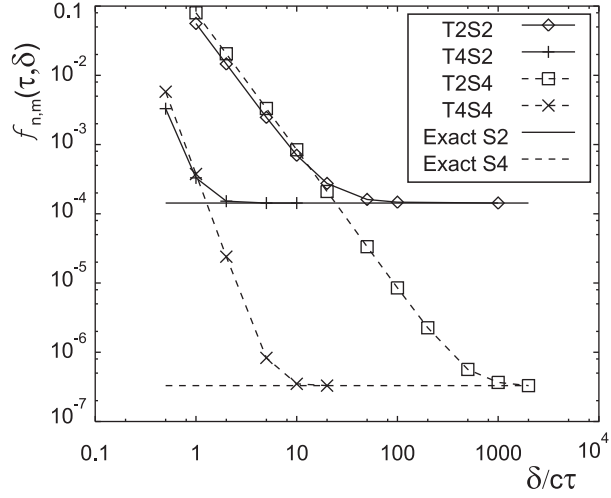}
\medskip
\caption{$f_{n,m}(\tau,\delta)$ as a function of $\delta/c\tau$
for the fixed mesh size $\delta=0.1$.
\label{fig:acc_time}}
\end{center}
\end{figure}

In Fig.~\ref{fig:rms_time} we plot $\Delta\mathbf{\Psi}_{n,m}(t)$ as a function
of the simulation time $t$ for fixed values of the mesh size $\delta$ and the
time step $\tau$ using both the T4S2 and the T4S4 algorithm.
We find that the error increases roughly proportional to the simulation
time:
\begin{equation}
\Delta\mathbf{\Psi}_{n,m}(t)\;=\;\sqrt{2}\,f_{n,m}(\tau,\delta)\,t\,,
\label{RMSerrorlin}
\end{equation}
where we used the prefactor $\sqrt{2}$ to ensure that
$0\leq f_{n,m}(\tau,\delta) \leq 1$.
The linear dependence of $\Delta\mathbf{\Psi}_{n,m}(t)$ on $t$ is clearly
visible only for the T4S2 algorithm but is also true for the T4S4 algorithm
with a much smaller slope $f_{4,4}(\tau,\delta)$.
Only at particular times $t$ when the wave packet hits the boundaries of
the cavity, the error $\Delta\mathbf{\Psi}_{4,4}(t)$ is seen to increase
nonlinearly in the time $t$ and takes a value that is of the same order
as $\Delta\mathbf{\Psi}_{4,2}(t)$.
This behavior, not described by Eq.~(\ref{RMSerrorlin}), is present in
fourth-order accurate spatial derivative implementations, in which the
calculation of the EM field components close to system boundaries refer to
several non-existing grid points.
To study the error $\Delta\mathbf{\Psi}_{n,m}(t)$ as a function of the
time step $\tau$ and the mesh size $\delta$, we compute
\begin{equation}
f_{n,m}(\tau,\delta)\;=\;\frac{1}{\sqrt{2}}
\frac{d}{dt}\Delta\mathbf{\Psi}_{n,m}(t) \,.
\end{equation}
In Fig.~\ref{fig:acc_time} we plot $f_{n,m}(\tau,\delta)$ as obtained for the
1D cavity by the four algorithms T2S2, T4S2, T2S4, and T4S4 as a function of
$\delta/c\tau$ for a fixed mesh size $\delta$.
For each algorithm T$n$S$m$ we find a linear decrease of
$\log[f_{n,m}(\tau,\delta)]$ with increasing values $\log[\delta/c\tau]$.
For the algorithms T4S2 and T4S4 we find that
$f_{4,m}(\tau,\delta)\propto \tau^4$,
while for the T2S2 and T2S4 algorithms
$f_{2,m}(\tau,\delta)\propto \tau^2$.
This numerical result is in agreement with the rigorous upper bound on the
error of the EM field vector Eq.~(\ref{psierrortau}).
For decreasing values of $\tau$, the error in the time integration becomes
negligible small and $f_{n,m}(\tau,\delta)$ reaches minimum values which
are indicated by the two lines `Exact S2' for the algorithms T$n$S2 and
`Exact S4' for the algorithms T$n$S4.
In fact, these two lines represent the numerical results that are obtained
for an exact time integration and $m$th-order accurate approximations to the
spatial derivatives.

Next, we study $f_{n,m}(\tau,\delta)$ as a function of the mesh size $\delta$
for the time step $\tau=0.1\delta/c$ to ensure that the accuracy of the time
integration remains constant.
\begin{figure}[h]
\begin{center}
\includegraphics[width=8.0cm]{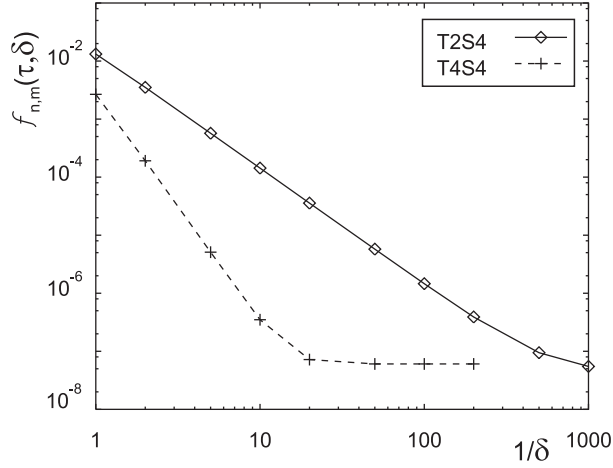}
\medskip
\caption{$f_{n,m}(\tau,\delta)$ as a function of $\delta$ for the time
step $\tau=0.1\delta/c$.
\label{fig:acc_mesh}}
\end{center}
\end{figure}
The numerical results are plotted in Fig.~\ref{fig:acc_mesh}.
We see that $\log[f_{n,m}(\tau,\delta)]$ decreases linearly with increasing
$\log[1/\delta]$ until it levels off.
At this point, the total number of operations has become so large that it
causes the numerical loss of accuracy.
Outside this regime we find for the T$n$S4 algorithms
$f_{n,4}(\tau,\delta)\propto\delta^4$ and for the T$n$S2 algorithms
$f_{n,2}(\tau,\delta)\propto\delta^2$.
In analogy to the upper bound Eq.~(\ref{psierrortau}), the upper bound for
the $m$th-order accurate approximation of the spatial derivatives is given by
\begin{equation}
\|\mathbf{\Psi}(t)\,-\,\mathbf{\Psi}_{n,m}(t)\|\;\leq\;
C_{n,m}\,t\,\delta^m\,,
\end{equation}
where $C_{n,m}$ is a constant.

\begin{figure}[h]
\begin{center}
\includegraphics[width=8.0cm]{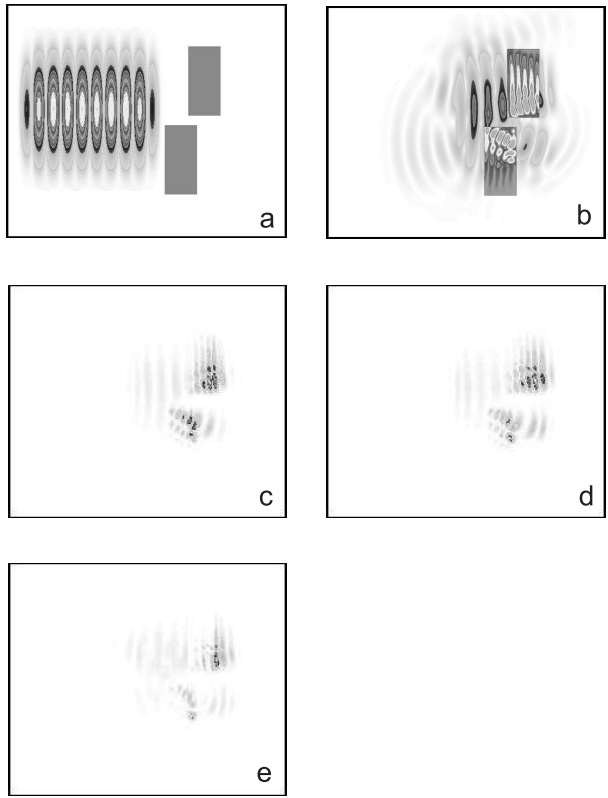}
\medskip
\caption{The error $\Delta w_{n,m}(\mathbf{r},t)$ for various mesh sizes and algorithms
in 2D.
(a) Initial energy density distribution $\mathbf{\Psi}(\mathbf{r}, t)^2$.
(b) Reference energy density distribution $\mathbf{\Psi}_{2,2}(\mathbf{r},t)^2$
at $t=6$ using the T2S2 algorithm with $\delta=0.025$.
(c) The error $\Delta w_{2,2}(\mathbf{r},t)$ on the energy density distribution
at $t=6$ using the T2S2 algorithm with $\delta=0.1$. The relative deviation is 26\%.
(d) The error $\Delta w_{2,2}(\mathbf{r},t)$ on the energy density distribution
at $t=6$ using the T2S2 algorithm with $\delta=0.05$. The relative deviation is 5.9\%.
(e) The error $\Delta w_{2,4}(\mathbf{r},t)$ on the energy density distribution
at $t=6$ using the T2S4 algorithm with $\delta=0.1$. The relative deviation is 5.1\%.
\label{fig:rmserr2d}}
\end{center}
\end{figure}

We consider a second example to illustrate the numerical performance of the
algorithms in 2D systems.
For the initial wave packet in the 2D cavity we make the ansatz
\begin{equation}
E_z(x,y,t)\;=\;\sin (k(x-x_0-ct))\exp
[-((x-x_0-ct)/\sigma_x)^{10}-((y-y_0)/\sigma_y)^2]\,.
\label{eqn:field_2d}
\end{equation}
At $t=0$ the wave packet is centered at $(x_0,y_0)$ and moves at $t>0$ with
velocity $c$ in the $x$-direction.
The energy of the wave packet is fixed by the wave number $k$ in the
oscillating factor and its envelope is Gaussian along the $y$-direction and
has sharp edges along the $x$-axis (due to the exponent $10$).
The 2D cavity of size $12\times 10$ with $\varepsilon=1$ and $\mu=1$ contains
two objects with dielectric constants $\varepsilon=5$ and $\mu=1$.
The parameters of the propagating wave packet are
$(\sigma_x,\sigma_y)=(1.66,1.29)$, $(x_0,y_0)=(3.5,5.5)$, and $k=5$.
In Fig.~\ref{fig:rmserr2d} we show the results for the error
Eq.~(\ref{errortemspa}) of the T2S2 and T2S4 algorithms with different mesh
sizes relative to a reference EM field vector $\mathbf{\Psi}(t)$ that is obtained
from the T2S2 algorithm at mesh size $\delta=0.025$.
In all simulations we kept $\tau=0.1\delta/c$ fixed to compare measurements of
constant accuracy in the time integration.
In Fig.~\ref{fig:rmserr2d} we show (a) the energy distribution of the initial
wave packet ($t=0$) and (b) the reference energy density distribution after
simulation time $t=6$ using the T2S2 algorithm.
In (c)-(e), the normalized spatial distribution of the error in the energy density
distribution,
\begin{equation}
\Delta w_{n,m}(\mathbf{r},t)\;=\;
|\mathbf{\Psi}(\mathbf{r},t)^2-\mathbf{\Psi}_{n,m}(\mathbf{r},t)^2|\,,
\end{equation}
is shown for, respectively, the algorithm
T2S2 with $\delta=0.1$, the algorithm T2S2 with $\delta=0.05$, and the algorithm
T2S4 with $\delta=0.1$.
We find that the improved spatial discretization implementation T2S4 with
$\delta=0.1$ performs as well as a poor man's implementation T2S2 with half the
mesh size.
The main advantage of using the T2S4 algorithm is that it used only $20\%$ of the
computer memory
%(2.2Mb versus 10Mb: about one fifth)
and $10\%$ of the CPU time
%(8.3s versus 113s: about one tenth) compared
with respect to the T2S2 algorithm.

%
%%%%%%%%%%%%%%%%%%%%%%%%%%%%%%%%%%%%%%%%%%%%%%%%%%%%%%%%%%%%%%%%%%%%%%%%%%%%
%
\section{Conclusions}\label{sec6}
%
%%%%%%%%%%%%%%%%%%%%%%%%%%%%%%%%%%%%%%%%%%%%%%%%%%%%%%%%%%%%%%%%%%%%%%%%%%%%
%

We have demonstrated that the previously introduced family of unconditionally
stable algorithms to solve the time-dependent Maxwell equations can be
implemented with a grid of variable mesh size and with a fourth-order accurate
approximation to the spatial derivatives.
The performance of the algorithms has been shown to increase significantly
as compared to the previously applied poor man's implementation while at
the same time their property of unconditional stability by construction
is preserved.
Performing numerical simulations on various physical model systems, we
found that a variable grid implementation can save orders of magnitude
in computer memory and CPU time for a physical system of unregular
geometrical shape or with strongly varying permeability and/or
permittivity.
Similar enhancements have been obtained for the fourth-order accurate
spatial derivative implementation which does not only reduce the numerical
dispersion but also improves the temporal and spatial accuracy of the
algorithms significantly.
Clearly, in close analogy to the implementation of the fourth-order
approximation of the spatial derivatives, the algorithms may be improved
by constructing higher-order approximations.
In general, we conclude that the family of unconditionally stable algorithms
does not only preserve the fundamental symmetries of the time-dependent Maxwell
equations but is also characterized by a high degree of flexibility that
allows to construct implementations that are required in different kinds
of specific applications.

%
%%%%%%%%%%%%%%%%%%%%%%%%%%%%%%%%%%%%%%%%%%%%%%%%%%%%%%%%%%%%%%%%%%%%%%%%%%%%
%
\section*{Acknowledgements}
%
%%%%%%%%%%%%%%%%%%%%%%%%%%%%%%%%%%%%%%%%%%%%%%%%%%%%%%%%%%%%%%%%%%%%%%%%%%%%
%
This work is partially supported by the Dutch `Stichting Nationale Computer
Faciliteiten' (NCF).
We thank W. Bruns for providing numerical results generated by the
program package GdfidL.

%%%%%%%%%%%%%%%%%%%%%%%%%%%%%%%%%%%%%%%%%%%%%%%%%%%%%%%%%%%%%%%%%%%%%%%%%%%%%
%%%%%%%%%%%%%%%%%%%%%%%%%%%%%%%%%%%%%%%%%%%%%%%%%%%%%%%%%%%%%%%%%%%%%%%%%%%%%
%%%%%%%%%%%%%%%%%%%%%%%%%%%%%%%%%%%%%%%%%%%%%%%%%%%%%%%%%%%%%%%%%%%%%%%%%%%%%

%
%
%
%
\end{document}